\documentclass[aps,pre,preprint,showpacs]{revtex4}
\usepackage{bm}
\usepackage{amsfonts}
\usepackage[dvips]{graphicx}
\usepackage{hyperref}
\hypersetup{ pdftitle={Mode coupling and conversion at
anticrossings treated via stationary perturbation technique},
pdfauthor={Dzmitry M. Shyroki}, pdfkeywords={} }
\begin{document}

\title{Mode coupling and conversion at anticrossings treated via
stationary perturbation technique}
\author{Dzmitry M. Shyroki}
\email{shyroki@tut.by} \affiliation{Department of Theoretical
Physics, Belarusian State University, Fr.~Skaryna avenue~4, Mensk
220080, Belarus}


\begin{abstract}
Intermodal interactions displayed through the phenomena of mode
coupling and conversion in optical systems are treated by means of
the Lindstedt-Poincar\'{e} perturbation method of strained
parameters more widely known in classical quantum mechanics and
quantum chemistry as the stationary perturbation technique. The
focus here is on the mode conversion at the points of virtual
phase matching (otherwise called anticrossings or avoided
crossings) associated with the maximum conversion efficiency. The
method is shown to provide a convenient tool to deal with
intermodal interactions at anticrossings --- interactions induced
by any kind of perturbation in dielectric index profile of the
waveguide, embracing optical inhomogeneity, magnetization of
arbitrary orientation, and nonlinearity. Closed-form analytic
expressions are derived for the minimum value of mode mismatch and
for the length of complete mode conversion (the coupling length,
or the beat length) in generic waveguiding systems exhibiting
anticrossings. Demonstrating the effectiveness of the method,
these general expressions are further applied to the case of
$\text{TE}_n \leftrightarrow \text{TM}_m$ mode conversion in (i) a
multilayer gyrotropic waveguide under piecewise-constant,
arbitrarily oriented magnetization, and (ii) an
optically-inhomogeneous planar dielectric waveguide
--- an example which the standard coupled-mode theory fails to
describe.
\end{abstract}

\maketitle

\section{\label{secInt}Introduction}
Mode coupling and conversion are commonly known and primarily
important phenomena in fiber and integrated optics, either
hindering or fueling the operation of numerous devices and
elements, e.g., electro-optical and acousto-optical switches and
modulators, waveguide couples, power splitters, wavelength
filters, and others~\cite{CoupledModeBOOKs}. When a mode of
certain configuration transverses the structure and converts into
some other mode, say, due to the externally induced perturbation
in optical properties or geometrical configuration of the
structure or due to the imperfectness of the materials used, or
when the exchange of energy between the guided modes of adjacent
waveguides occurs
--- anyway we face the phenomena of mode coupling and conversion.
The crucial regime for the mode coupling is in the vicinity of the
points of virtual phase matching (otherwise referred to as
``anticrossings'' or ``avoided crossings'' --- a terminology
brought about from quantum mechanics and solid-state physics). At
anticrossings, the complete mode conversion can be achieved, which
makes tailoring this regime particularly important for the
applications. For a positive example, let me mention a promising
method for controllable dispersion compensation in photonic
bandgap fibers based on weak interactions at anticrossings between
the core-guided mode and a mode localized in an intentionally
introduced defect of the crystal~\cite{DispersionTailoring}. An
opposing example of technology strains due to the undesirable
intermodal interactions is radiation losses caused by the coupling
of the guided mode to radiation modes in a two-dimensional
photonic crystal etched into a planar
waveguide~\cite{RadiationLossesBOOK}.

A dominating theoretical tool for the whole of mode coupling and
conversion phenomena (except a few cases when rigorous analytical
treatment is possible~\cite{RigorousTheory}) is the coupled-mode,
or the coupled-wave, formalism~\cite{CoupledModeBOOKs,Huang94}
that was first proposed more than fifty years ago and since then
became commonplace in both optical engineering research and
textbook literature. The mathematical foundation of all the
modifications of the coupled-mode formalism is the method of
variation of independent coefficients --- one of the fruitful
methods in perturbation theory~\cite{PerturbationMethodsBOOK}.
Nevertheless, the formalism grounds upon the differing amount of
approximations
--- e.g., that of slowly-varying dielectric function
$\epsilon(\mathbf{r})$, so that the electric field
$\mathbf{E}(\mathbf{r})$ is assumed to satisfy
$\bm\nabla(\bm\nabla\cdot\mathbf E)=\mathbf 0$; the so-called
parabolic approximation; and others --- which certainly limits the
usefulness of the model. Moreover, the resultant ``coupled modes''
often fail to satisfy orthogonality relations and boundary
conditions in the actual (perturbed) structure. Finally, the
coupled-mode formalism ignores all of the explicit mathematical
parallels to the classical quantum-mechanics problems that also
exploit perturbation techniques, in particular the widely known
perturbation method for stationary
levels~\cite{WilcoxBOOK,LLQuantumMechanicsBOOK} --- a modification
of the Lindstedt-Poincar\'{e} method of strained parameters
sometimes called the Rayleigh-Schr\"{o}dinger method (see
Refs.~\cite{PerturbationMethodsBOOK,WilcoxBOOK} for the
bibliographical details). Meanwhile, establishing and tracking the
analogies between classical quantum mechanics and optics tends to
be a really stimulating approach --- both through adopting the
formalism of electromagnetic-wave propagation in dielectric
materials to collisionless  propagation of effective-mass electron
waves in semiconductor crystals~\cite{Henderson92-93}, and on the
other hand through applying the formalism of quantum-mechanical
solid-state electronics to electromagnetic propagation in periodic
dielectric structures which proved to be a very successful
programme during the last decade~\cite{PhC,PhCBOOK}.

In this article, I translate the classical stationary perturbation
technique --- i.e., the method of strained parameters --- to the
case of mode coupling and conversion in optical waveguiding
systems. The problem of source-free light propagation is
formulated in \hyperref[sec2]{Sec.~\ref*{sec2}} as an ordinary
eigenvalue problem in the squared free-space wavenumber
$\kappa\equiv k^2=\omega^2/c^2$ (or, actually, in the angular
frequency $\omega$) for the magnetic field $\mathbf H(\mathbf r)$,
and perturbations in dielectric index profile of the waveguide are
considered, embracing the cases of optical inhomogeneity,
magnetization of arbitrary orientation, and nonlinearity. Both the
eigenvalues and the eigenvectors are expanded into the series in
perturbation parameter, giving rise to an electromagnetic
counterpart of the quantum-mechanics stationary perturbation
method as an alternative to the coupled-mode formalism. Although
physical interpretation of the former in the case of nondegenerate
spectrum of eigenvalues looks rather questionable from the point
of experiment, the method however provides promptly solvable and
naturally interpretable treatment of the modal behavior at the
anticrossings, which is an issue in the mode conversion analysis.
To demonstrate the efficiency of the method (to distinguish from
the coupled-mode theory, the qualifying term ``stationary'' is
retained here with the primary meaning that the coefficients of an
expansion into which an input wave is decomposed are not varying
with coordinates), I refer in \hyperref[sec3]{Sec.~\ref*{sec3}} to
the two practice-targeted examples from integrated optics
concerning $\text{TE} \leftrightarrow \text{TM}$ mode coupling: a
multilayer gyrotropic waveguide subject to constant, arbitrarily
oriented magnetic field, and an optically-inhomogeneous planar
dielectric waveguide
--- an example which falls out of the scope of the standard
coulped-mode theory. The unperturbed basis for the both cases is a
multilayer waveguide composed of linear isotropic dielectric
materials --- a simple structure known to exhibit under certain
circumstances the perfect phase matching of guided TE and TM
modes~\cite{ShyrokiLavrinenko03}, thus it is natural to implement
here the developed theory for the mode conversion at
anticrossings. For both systems, I pursue the formulated way to
derive closed-form first-order analytic expressions for the
minimum value of mode mismatch and hence for the length of
complete $\text{TE}_n \leftrightarrow \text{TM}_m$ mode conversion
(the coupling length, or the beat length). Finally, I conclude in
\hyperref[sec4]{Sec.~\ref*{sec4}} with some claims concerning
further possible applications and modifications of the stationary
perturbation technique for the optical waveguiding theory.

\section{\label{sec2}H-eigenproblem and anticrossings}
\subsection{H-eigenproblem}
Let me start with the wave equation for the magnetic field
\begin{equation} \label{H-eigenproblem}
\bm\nabla\times\left(\epsilon^{-1} \bm\nabla\times\mathbf H
\right) = \kappa \mathbf H,
\end{equation}
where $\epsilon=\epsilon(\mathbf r)$ is the dielectric
permittivity distribution, $\kappa \equiv k^2 = \omega^{2}/c^2$
the squared free-space wave number, $\omega$ the angular
frequency, $c$ the vacuum speed of light, and the field $\mathbf
H$ satisfies additionally
\begin{equation} \label{divH}
\bm\nabla\cdot\mathbf H=0.
\end{equation}
Eq.~(\ref{H-eigenproblem}) can be treated as an ordinary
eigenvalue problem in $\kappa$ (or, actually, in $\omega$) for the
field $\mathbf H$, with Maxwellian operator $\mathcal{M}$ defined
by
\begin{equation} \label{Maxwellian}
\mathcal{M}\mathbf H \equiv \bm\nabla\times\left(\epsilon^{-1}
\bm\nabla\times\mathbf H \right).
\end{equation}
In fact, a similar equation can be written for, say, the $\mathbf
E$ field:
\begin{equation} \label{E-eigenproblem}
\epsilon^{-1} \bm\nabla\times\left(\bm\nabla\times\mathbf E
\right)= \kappa \mathbf E,
\end{equation}
but there are two sound reasons to restrict oneself to the
$\mathbf H$-eigenproblem (\ref{H-eigenproblem}), but not to its
$\mathbf E$-counterpart (\ref{E-eigenproblem}): first, an
accompanying to Eq.~(\ref{H-eigenproblem}) divergence equation
(\ref{divH}) is clearly simpler than $\bm\nabla\cdot(\epsilon
\mathbf E)=0$ --- a satellite equation for the $\mathbf
E$-eigenproblem; second, given the dielectric index a real scalar
function of coordinates --- i.e., a real scalar field
--- or, more generally, a Hermitian dyadic field, the Maxwellian
(\ref{Maxwellian}) becomes Hermitian too~\cite{LLContiMediaBOOK}
--- the fact that is though not crucial in macroscopic
electrodynamics (non-Hermitian Maxwellian would generete a set of
nonorthogonal eigenmodes and complex-valued eigenfrequencies --- a
situation normally faced in optics of lossy qyrotropic
media~\cite{FedorovBOOK}) nor pertinent to the present treatment,
but aesthetically pleasant and legitimates many of the
cross-references between Eq.~(\ref{H-eigenproblem}) and the
stationary Schr\"{o}dinger eigenproblem with Hermitian Hamiltonian
as well.

For a large class of waveguiding systems, namely, for those
exhibiting continuous translation symmetry along the direction of
light propagation (say, the $z$ direction) we can assign harmonic
dependence of the fields on that direction; in particular, the
magnetic field reads
\begin{equation} \label{ExpBeta}
\mathbf H(\mathbf r;\beta) = e^{i\beta z}\bm{\mathcal{H}}(x,y),
\end{equation}
where $\beta$ is the propagation constant, so that for a given
$z$-independent dielectric index profile $\epsilon_{(0)}(x,y)$
Eq.~(\ref{H-eigenproblem}) reduces to an eigenvalue problem for
$\bm{\mathcal{H}}(x,y)$; the reduced eigenproblem operator depends
parametrically on $\beta$ then and at a fixed $\beta$ spawns a set
of eigenvalues $\kappa(\beta)$, thus yielding dispersion structure
$\mathcal{D}(\kappa,\beta)=0$ of a perfect waveguide. If an actual
dielectric permittivity distribution $\epsilon(\mathbf r)$ of the
waveguide happens to differ from $\epsilon_{(0)}(x,y)$ for which
Eq.~(\ref{ExpBeta}) holds, that is $\epsilon(\mathbf r) =
\epsilon_{(0)}(x,y) + \varepsilon\delta\epsilon(\mathbf r)$ (we do
not need to specify at this stage whether $\delta\epsilon(\mathbf
r)$ is a scalar or a tensor field) and
\begin{equation}
\epsilon^{-1}(\mathbf r) = \xi_{(0)}(x,y) +
\varepsilon\xi_{(1)}(\mathbf r) + \varepsilon^2\xi_{(2)}(\mathbf
r) + \dots,
\end{equation}
where $\xi_{(i)}(\mathbf r) = [-\delta\epsilon(\mathbf
r)]^i/[\epsilon_{(0)}(x,y)]^{i+1}$, $i=0,1,\dots$, then in the
spirit of the Lindstedt-Poincar\'{e} perturbation method of
strained parameters~\cite[ch.~3]{PerturbationMethodsBOOK} one can
rewrite Eq.~(\ref{H-eigenproblem}) as
\begin{eqnarray} \label{H-eigenproblem-perturbed}
(\mathcal{M}^{(0)} + \varepsilon\mathcal{M}^{(1)} + \varepsilon^2
\mathcal{M}^{(2)} + \dots)\mathbf H = \kappa \mathbf H,
\\
\mathcal{M}^{(i)}\mathbf H \equiv
\bm\nabla\times(\xi_{(i)}\bm\nabla\times\mathbf H),
\end{eqnarray}
in terms of the unperturbed Maxwellian $\mathcal{M}^{(0)}$ and the
higher-order perturbation operators, and expand then the
eigenvectors $\mathbf H_n$ of the perturbed problem into a series
in $\varepsilon$:
\begin{equation} \label{H-expansion}
\mathbf H_n = \mathbf H^{(0)}_n + \varepsilon\mathbf H^{(1)}_n +
\varepsilon^2\mathbf H^{(2)}_n + O(\varepsilon^2),
\end{equation}
and similarly for the eigenvalues:
\begin{equation} \label{k-expansion}
\kappa_n = \kappa^{(0)}_n + \varepsilon \kappa^{(1)}_n +
\varepsilon^2 \kappa^{(2)}_n + O(\varepsilon^2).
\end{equation}
The procedure for further solving
Eq.~(\ref{H-eigenproblem-perturbed}) using expansions
(\ref{H-expansion}) and (\ref{k-expansion}), i.e., for finding the
unknown functions $\mathbf H^{(i)}_n$ and $\kappa^{(i)}_n$, is
quite well-developed in perturbation theory. Substituting
Eqs.~(\ref{H-expansion}), (\ref{k-expansion}) into
Eq.~(\ref{H-eigenproblem-perturbed}) yields for the zeroth-order
and linear in $\varepsilon$ terms:
\begin{eqnarray}
\mathcal{M}^{(0)}\mathbf H^{(0)}_n &=& \kappa^{(0)}_n\mathbf
H^{(0)}_n, \label{zeroth}
\\
\mathcal{M}^{(0)}\mathbf H^{(1)}_n + \mathcal{M}^{(1)}\mathbf
H^{(0)}_n &=& \kappa^{(0)}_n\mathbf H^{(1)}_n +
\kappa^{(1)}_n\mathbf H^{(0)}_n, \label{first}
\end{eqnarray}
with the homogeneous boundary conditions $\bm\nabla\cdot\mathbf
H^{(i)}_n=0$. The solution to the unperturbed problem
(\ref{zeroth}) is assumed to be known; to say more --- and this is
an important point --- I further assume all the $\mathbf
H^{(0)}_n$ vectors to be of the form (\ref{ExpBeta}) which is
likely to embrace all the cases of practical interest. The
subsequent steps of the method depend essentially on whether the
spectrum of $\mathcal{M}^{(0)}$ is degenerate or not.

\subsection{Nondegenerate spectrum}
If there are no degenerate eigenvalues among those of
$\mathcal{M}^{(0)}$ spectrum, then we customarily obtain up to the
first-order correcting terms~\cite{PerturbationMethodsBOOK}:
\begin{equation} \label{H-expansion-nondeg}
\mathbf H_n(\mathbf r;\beta) \approx \mathbf H^{(0)}_n(\mathbf
r;\beta) + \varepsilon\sum_{m\neq
n}\frac{\mathcal{M}^{(1)}_{nm}(\beta)}{\kappa^{(0)}_n -
\kappa^{(0)}_m}\mathbf H^{(0)}_m(\mathbf r;\beta)
\end{equation}
and
\begin{equation} \label{k-expansion-nondeg}
\kappa_n(\beta) \approx \kappa^{(0)}_n(\beta) + \varepsilon
\mathcal{M}^{(1)}_{nn}(\beta),
\end{equation}
where the first-order coupling matrix
\begin{equation}\label{M1}
\mathcal{M}^{(1)}_{nm}(\beta) \equiv \mathcal{N}^{-1}\!\!\int_W
\mathbf H^{*(0)}_m(\mathbf r;\beta)\,\mathcal{M}^{(1)}(\mathbf
r)\,\mathbf H^{(0)}_n(\mathbf r;\beta)\,dv.
\end{equation}
The integration volume $W$ embracing the waveguide is formally
infinite, $dv=dx\,dy\,dz$, the normalizing factor $\mathcal{N}$
equals either $Z$ --- the distance of light propagation within a
waveguide in the $z$ direction (i.e., the length of the waveguide)
--- in the case of cylindrical (two-dimensional) geometry, or
$S = YZ$ --- the $yz$ square of a waveguiding cell --- for planar
(one-dimensional) systems. This normalization is due to specific
orthogonality condition assigned to the $\mathbf H^{(0)}_n$ modes:
\begin{eqnarray}
\int_W \mathbf H^{(0)}_n\mathbf H^{*(0)}_m\,dv &=&
\delta_{nm}\int_W
\left|\mathbf H^{(0)}_m\right|^2dv \nonumber\\
&=& \left\{ \begin{array}{ll}
\delta_{nm}Z\int\left|\bm{\mathcal{H}}_m(x,y)\right|^2dx\,dy
=\delta_{nm}Z &
\text{(cylindrical waveguide),}\\
\delta_{nm}S\int\left|\bm{\mathcal{H}}_m(x)\right|^2dx=\delta_{nm}S
& \text{(planar waveguide).}
\end{array}\right.
\end{eqnarray}

\begin{figure}
\includegraphics{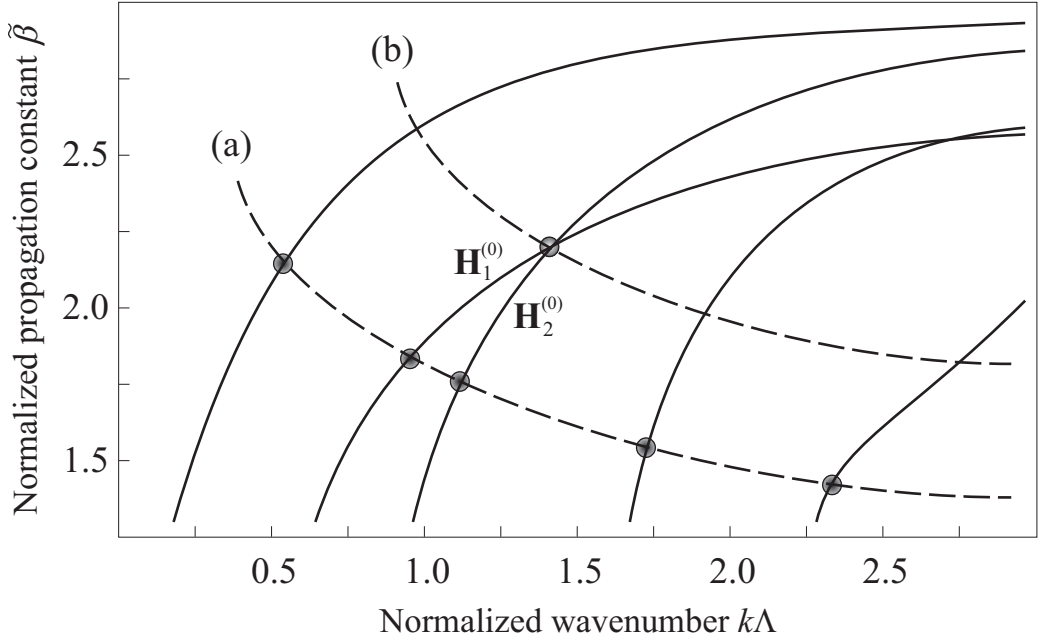}
\caption{Modal dispersion curves (--------) of the unperturbed
hypothetical structure and $\beta = \text{const}$ lines (-- -- --)
for the case of (a) nondegenerate spectrum of $\mathcal{M}^{(0)}$
and (b) the two modes, $\mathbf H_1^{(0)}$ and $\mathbf
H_2^{(0)}$, exhibiting degeneration. Here $\Lambda$ is a
characteristic dimension of the structure; the grey circles mark
the modes related to the expansions (\ref{H-expansion-nondeg}) and
(\ref{H-expansion-deg}).}\label{f1}
\end{figure}

I explicitly designated the (parametric) dependence of all the
quantities in Eqs.~(\ref{H-expansion-nondeg}),
(\ref{k-expansion-nondeg}), and (\ref{M1}) on $\beta$ entailed by
the previous assumption that the unperturbed modes have the form
as per Eq. (\ref{ExpBeta}). Now, the restriction $\beta =
\text{fixed}$ for the set of nondegenerate modes $\mathbf
H^{(0)}_n e^{-i\omega_{n}t}\propto e^{i(\beta z - \omega_{n}t)} =
e^{ik_{n}(\tilde\beta_{n}z - ct)}$ into which the \textit{n}th
perturbed eigenmode $\mathbf H_n$ is expanded via
Eq.~(\ref{H-expansion-nondeg}) --- modes characterized thus by
essentially different eigenwavenumbers $k_n$ and differing
normalized propagation constants $\tilde\beta_{n} \equiv
\beta/k_n$ --- seems to correspond to an utterly odd situation
from the point of experiment (see \hyperref[f1]{Fig.~\ref*{f1}}).
To aid this shortcoming somehow, one might consider
Eq.~(\ref{H-eigenproblem}) as a generalized eigenproblem in, e.g.,
$\beta$ or $\tilde\beta$, parametrically dependent on $\kappa$,
which would lead to expansions like (\ref{H-expansion-nondeg}) but
into the $\{\mathbf H^{(0)}_n(\mathbf r; \kappa)\}$ set. Of
course, this severely complicates the formalism --- and probably
here is an issue why the coupled-mode theory has been exclusively
dominating for decades in the field. Fortunately, the case of
quasi-degenerate eigenvalues allows promptly solvable and easily
interpretable treatment in terms of the ordinary $\mathbf
H$-eigenproblem in $\kappa$.

\subsection{Degenerate spectrum and anticrossings}
Let us focus on the case of two-degenerate or nearly degenerate
eigenmodes $\mathbf H^{(0)}_1$ and $\mathbf H^{(0)}_2$. Borrowing
the known result from quantum
mechanics~\cite{LLQuantumMechanicsBOOK}, we see that in this case
both the $\mathbf H^{(0)}_1$ and $\mathbf H^{(0)}_2$ modes
dominate the expansion (\ref{H-expansion}) of the related
fundamental modes
\begin{equation}\label{H-expansion-deg}
\mathbf H_{\pm} = C_1^{\pm}\mathbf H^{(0)}_1 + C_2^{\pm}\mathbf
H^{(0)}_2 + O(\varepsilon)
\end{equation}
with the constant (in particular, $z$-independent) coefficients
\begin{widetext}
\begin{equation} \label{C1}
C^{\pm}_1 = \left[
\frac{\mathcal{M}_{12}}{2\sqrt{\mathcal{M}_{12}\mathcal{M}_{21}}}
\left( 1\pm\frac{\mathcal{M}_{11} -
\mathcal{M}_{22}}{\sqrt{(\mathcal{M}_{11} - \mathcal{M}_{22})^2 +
4\mathcal{M}_{12}\mathcal{M}_{21}}} \right) \right]^{1\over 2},
\end{equation}
\begin{equation} \label{C2}
C^{\pm}_2 = \pm\left[
\frac{\mathcal{M}_{12}}{2\sqrt{\mathcal{M}_{12}\mathcal{M}_{21}}}
\left( 1\mp\frac{\mathcal{M}_{11} -
\mathcal{M}_{22}}{\sqrt{(\mathcal{M}_{11} - \mathcal{M}_{22})^2 +
4\mathcal{M}_{12}\mathcal{M}_{21}}} \right) \right]^{1\over 2}.
\end{equation}
Here and below in the article I pursue only the first-order
approximations and hence omit the overly superscript $^{(1)}$ over
the coupling matrix elements $\mathcal{M}_{nm}$, now
$\beta$-independent, defined by Eq.~(\ref{M1}). The first-order
correction to the corresponding eigenvalues leads
to~\cite{LLQuantumMechanicsBOOK}
\begin{equation}\label{k-deg}
\kappa_{\pm} = \frac{\kappa^{(0)}_1 + \kappa^{(0)}_2 +
\varepsilon( \mathcal{M}_{11} + \mathcal{M}_{22} )}{2} \pm
\sqrt{\left[ \frac{\kappa^{(0)}_1 - \kappa^{(0)}_2 +
\varepsilon(\mathcal{M}_{11} - \mathcal{M}_{22})}{2} \right]^2+
\varepsilon^2\mathcal{M}_{12}\mathcal{M}_{21}}.
\end{equation}
\end{widetext}
In quantum mechanics, this situation corresponds to
quasi-degenerate energy levels of a quantum system and is
associated with enormous variety of effects and treatments ---
take the Landau-Zener model for example~\cite{LZmodel}, nourishing
nearly 100 publications a year. In the realm of optics,
Eq.~(\ref{H-expansion-deg}) with coefficients given by
Eqs.~(\ref{C1}), (\ref{C2}) is fully consistent with the general
statement that complete mode conversion can be achieved in any
system exhibiting anticrossings. Indeed, instead of the $\mathbf
H^{(0)}_1$ mode and the $\mathbf H^{(0)}_2$ mode crossing one
another on the dispersion structure diagram
(\hyperref[f1]{Fig.~\ref*{f1}}), at the anticrossings the $\mathbf
H^{(0)}_1$ mode ``transforms'' continuously into the $\mathbf
H^{(0)}_2$ one, and vice-versa, over a certain frequency range, so
it is intuitive to expect that somewhere in the vicinity of the
virtual crossing point the $\mathbf H^{(0)}_1$ and $\mathbf
H^{(0)}_2$ modes would be equally unsuitable to play the role of
the fundamental modes $\mathbf H_{+}$ and $\mathbf H_{-}$, and if
the original (input) wave is $\mathbf H^{(0)}_1$ or $\mathbf
H^{(0)}_2$, it wouldn't ``feel'' which of the fundamental modes,
$\mathbf H_{+}$ or $\mathbf H_{-}$, is more suited for it, and
thus would oscillate between $\mathbf H^{(0)}_1$ and $\mathbf
H^{(0)}_2$. Mathematically, these heuristic speculations are
expressed via the requirement
\begin{equation}
|C_1^{+}| = |C_1^{-}| = |C_2^{+}| = |C_2^{-}| = \frac{1}{\sqrt 2}
\end{equation}
(under appropriate normalization) necessary for the complete mode
conversion to occur, which is equivalent, as immediately follows
from the comparison of Eqs.~(\ref{C1}) and (\ref{C2}), to
canceling out the $\mathcal{M}_{11}$ and $\mathcal{M}_{22}$
elements under the root in Eqs.~(\ref{C1}), (\ref{C2}), and
(\ref{k-deg}) due to
\begin{equation}\label{M11=M22}
\mathcal{M}_{11} = \mathcal{M}_{22}.
\end{equation}
This condition does not necessarily correspond to the case of
stringently degenerate eigenvalues $\kappa^{(0)}_1$ and
$\kappa^{(0)}_2$, but with a good deal of reason we may assume
$\kappa^{(0)}_1 - \kappa^{(0)}_2 = O(\varepsilon^2)$ to estimate
the minimum mode mismatch $\Delta k_{min}$ from Eq.~(\ref{k-deg}).
With this assumption, it is under condition (\ref{M11=M22}) that
the quantity
\begin{equation}
\Delta\kappa_{min} = 2k_0\Delta k_{min},
\end{equation}
where $k_0=\omega_0/c$, $\omega_0$ is the frequency of a virtual
mode crossing (phase matching), reaches its minimum value of
$2\varepsilon\sqrt{\mathcal{M}_{12}\mathcal{M}_{21}}$, thus the
minimum eigenmode mismatch $\Delta k_{min}$ along the
$\beta=\text{const}$ curve corresponds to the complete mode
conversion and reads
\begin{equation} \label{MinModeMismatch-gen}
\Delta k_{min} = \varepsilon
k_0^{-1}\sqrt{\mathcal{M}_{12}\mathcal{M}_{21}}.
\end{equation}
In a nonabsorbing waveguide (in Sec.~\ref{sec3} we will deal with
the two) $\mathcal{M}_{nm}$ is a Hermitian matrix, hence
$\mathcal{M}_{12}=\mathcal{M}_{21}^*$ and
Eq.~(\ref{MinModeMismatch-gen}) is simplified to
\begin{equation}\label{MinModeMismatch}
\Delta k_{min} = \varepsilon k_0^{-1}|\mathcal{M}_{12}|.
\end{equation}
Complete mode conversion occurs when the $\mathbf H^{(0)}_1$ and
$\mathbf H^{(0)}_2$ modes propagating in the $z$ direction
accumulate a phase difference of $\pi$: $\Delta k_{min}\tilde\beta
z_c = \pi$; substituting here Eq.~(\ref{MinModeMismatch}), we
obtain for the corresponding coupling length:
\begin{equation}\label{CouplingLength}
z_c = \frac{\pi k_0}{\varepsilon\tilde\beta|\mathcal{M}_{12}|},
\end{equation}
where, as before [see Eq.~(\ref{M1})], the coupling matrix element
\begin{equation}\label{M12}
\mathcal{M}_{12} = \mathcal{N}^{-1}\!\!\int_W
[\bm\nabla\times(\xi_{(1)}\bm\nabla\times\mathbf
H^{(0)}_1)]\cdot\mathbf H^{*(0)}_2 dv.
\end{equation}
In contrast to the coupled-mode theory, the $\mathcal{M}_{12}$
element might and, in general, does depend on the length of a
waveguide along the $z$ direction because of the possible $z$
dependence of the $\xi_{(1)}$ function. In what follows I
calculate $\mathcal{M}_{12}$ and hence $z_c$ explicitly for the
two specific systems relevant to integrated optics: for a
multilayer gyrotropic stack and for an optically-inhomogeneous
waveguide.

\section{\label{sec3}Mode conversion at planar geometry}
\subsection{Gyrotropic waveguide}
Let the unperturbed waveguide be described by the scalar
dielectric permittivity distribution $\epsilon_{(0)}(\mathbf r)$.
As argued by symmetry reasoning, relativity considerations, and
energy conservation~\cite{LLContiMediaBOOK,FedorovBOOK}, the most
general form to allow for the electromagnetic disturbance induced
by the $\varepsilon \mathbf h$ field applied to (initially)
isotropic medium is
\begin{equation}
\tensor\epsilon = \epsilon_{(0)}\tensor{\mathsf I} +
i\varepsilon\zeta\mathbf h^\times + \varepsilon^2\eta(\mathbf
h\otimes\mathbf h - \mathbf h^2),
\end{equation}
where $\tensor{\mathsf I}$ is a unit three-dimensional dyadic;
$\zeta$ and $\eta$ are real functions of $\mathbf h^2$; the
superscript $^\times$ over the $\mathbf h$ vector denotes the
antisymmetric dyadic dual to $\mathbf h$, so that $\mathbf
h^\times\mathbf g=\mathbf h\times\mathbf g$ gives conventional
vector product of $\mathbf h$ and $\mathbf g$; the circled cross
$\otimes$ denotes the outer product --- the dyad: $\mathbf
h\otimes\mathbf g=h_i g_j$, $i,j=1,2,3$. If $\varepsilon$ is
small, we may assume $\zeta$ and $\eta$ to be constant (this
assumption leads to the introduction of gyration
vector~\cite{LLContiMediaBOOK,FedorovBOOK}) and write down for the
inverse to the $\tensor\epsilon$ dyadic~\cite{FedorovBOOK}
\begin{equation}
\tensor\epsilon^{-1} = \epsilon_{(0)}^{-1}\tensor{\mathsf I} -
i\varepsilon \mathbf u^\times + O(\varepsilon^2),
\end{equation}
in terms of $\mathbf u \equiv \epsilon_{(0)}^{-2}\zeta\mathbf h$,
so that the coupling matrix element (\ref{M12}) can be written as
\begin{equation} \label{M12gyr}
\mathcal{M}_{12}^{gyr} = - kS^{-1}\!\!\int_W
[\bm\nabla\times(\mathbf u\times\mathbf D^{(0)}_1)]\cdot\mathbf
H^{*(0)}_2 dv,
\end{equation}
where I used the frequency-domain Maxwell's equation for
$\bm\nabla \times \mathbf H$ in the source-free region, $\bm\nabla
\times \mathbf H = - ik\mathbf D$. Equation (\ref{M12gyr}) can be
significantly simplified by applying the vector
identity~\cite{KornKornBOOK}
\begin{equation} \label{VectorIdentity}
\bm\nabla\times(\mathbf u\times\mathbf D) = \mathbf u(\bm\nabla
\cdot\mathbf D) - \mathbf D(\bm\nabla \cdot\mathbf u) + (\mathbf D
\cdot \bm\nabla)\mathbf u - (\mathbf u \cdot \bm\nabla)\mathbf D
\end{equation}
valid for any three-vectors $\mathbf u$ and $\mathbf D$. The first
term on the right vanishes immediately due to $\bm\nabla \cdot
\mathbf D=0$; then, in a specific but physically reasonable and
normally considered case of $\mathbf u$ being a constant (or
piecewise-constant) vector in a given volume $V$ and equaling zero
outside of it, the second and third terms of the sum in
Eq.~(\ref{VectorIdentity}) being multiplied by $\mathbf H$ and
integrated as per Eq.~(\ref{M12gyr}) lead to the surface integrals
instead of the volume ones. For the second term one obtains:
$\bm\nabla\cdot\mathbf u = - (\mathbf{\hat n}\cdot\mathbf
u)\,\delta(\mathbf r - \mathbf R)$, where $\mathbf R= \mathbf
R(\mathbf r)$ defines the boundary $\Sigma$ of $V$, $\mathbf{\hat
n}$ is the \textit{outer} normal to $\Sigma$; hence
\begin{equation}\label{SecondTerm}
\int_W [\mathbf D(\bm\nabla\cdot\mathbf u)]\cdot\mathbf H \, dv =
- \int_\Sigma [\mathbf D(\mathbf{\hat n}\cdot\mathbf
u)]\cdot\mathbf H \, d\sigma.
\end{equation}
For the third term we similarly have: $(\mathbf D \cdot
\bm\nabla)\mathbf u = - (\mathbf D \cdot \mathbf{\hat n})\mathbf u
\, \delta(\mathbf r - \mathbf R)$, hence for an integral
\begin{equation}\label{ThirdTerm}
\int_W [(\mathbf D \cdot \bm\nabla)\mathbf u] \cdot\mathbf H  \,dv
= -\int_\Sigma [(\mathbf D \cdot \mathbf{\hat n})\mathbf u]
\cdot\mathbf H \, d\sigma.
\end{equation}
Thus for the case of constantly magnetized waveguide (up to this
point, no assumptions have been made regarding its actual
geometry) the integral in Eq.~(\ref{M12gyr}) is reduced via
Eqs.~(\ref{VectorIdentity}), (\ref{SecondTerm}), and
(\ref{ThirdTerm}) to
\begin{equation} \label{ThreeIntegrals}
\int_W [\bm\nabla\times(\mathbf u\times\mathbf D]\cdot\mathbf H \,
dv = \int_\Sigma [\mathbf D(\mathbf{\hat n}\cdot\mathbf
u)]\cdot\mathbf H \, d\sigma - \int_\Sigma [(\mathbf D \cdot
\mathbf{\hat n})\mathbf u] \cdot\mathbf H \, d\sigma - \int_V
[(\mathbf u \cdot \bm\nabla)\mathbf D] \cdot\mathbf H \,dv.
\end{equation}

Now let us switch to the planar structures. In a planar
unperturbed waveguide the eigenmodes are classified in terms of
TE, or \textit{s}-polarized, and TM, or \textit{p}-polarized waves
(for an elegant derivation of this common fact via symmetry
considerations see Ref.~\cite{PhCBOOK}). In a coordinate system in
which the $x$ axes is normal to the bimedium interfaces and the
light energy is guided in the $z$ direction, we can write down, to
evaluate the $\mathcal{M}_{sp}^{gyr}$ element as per
Eq.~(\ref{M12gyr}):
\begin{equation} \label{Ds}
\mathbf D_1^{(0)}=\mathbf D_{s} = e^{i\beta z}\phi_{s}(x)
\mathbf{\hat y}
\end{equation}
for the $s$-polarized mode, and
\begin{equation} \label{Hp}
\mathbf H_2^{(0)}=\mathbf H_{p} = e^{i\beta z}\psi_{p}(x)
\mathbf{\hat y}
\end{equation}
for the $p$-polarized one. Here $\phi_{s}(x)$ and $\psi_{p}(x)$
are the lateral distributions of $\mathbf D_{s}$ and $\mathbf
H_{p}$ respectively, obtained from the unperturbed Maxwell's
equations for the given virtually-phase-matched modes. We see
immediately from Eq.~(\ref{Ds}) that the second surface integral
in Eq.~(\ref{ThreeIntegrals}) vanishes at planar geometry, since
$\mathbf D_{s}\cdot\mathbf{\hat n} = \mathbf
D_{s}\cdot\mathbf{\hat x} = \mathbf D_{s}\cdot(-\mathbf{\hat x}) =
0$ holds identically in this case. Finally we arrive at
\begin{eqnarray} \label{M12gyr-planar}
\mathcal{M}_{sp}^{gyr} &=& - kS^{-1}\!\!\int_\Sigma [\mathbf
D_s(\mathbf{\hat n}\cdot\mathbf u)]\cdot\mathbf H^{*}_p \, d\sigma
+ kS^{-1}\!\!\int_V [(\mathbf u \cdot \bm\nabla)\mathbf D_s]\cdot\mathbf H^{*}_p \,dv \nonumber\\
&=& - kS^{-1}\!\!\int_V u_x\left(\frac{d\phi_s}{dx}\psi_p +
\phi_s\frac{d\psi_p}{dx}\right)\,dv
+ kS^{-1}\!\!\int_V \left(u_x\frac{d\phi_s}{dx}\psi_p + i\beta u_z\phi_s\psi_p\right)\,dv \nonumber\\
&=& -k\left[u_x^\alpha
J^\alpha\left(\frac{\phi_s\,d\psi_p}{dx}\right) - iu_z^\alpha
J^\alpha\left(\frac{\phi_s\psi_p}{\beta^{-1}}\right) \right],
\end{eqnarray}
where $\mathbf u^\alpha$ is the value of vector $\mathbf u$ in the
$\alpha$th layer bounded between the $x = x_{\alpha-1}$ and $x =
x_\alpha$ planes, and the integrals
\begin{eqnarray}
J^\alpha\left(\frac{\phi_s\,d\psi_p}{dx}\right)
&=&\int_{x_{\alpha-1}}^{x_\alpha} \frac{\phi_s(x)\,d\psi_p(x)}{dx}
\,dx,\label{Jdx}
\\
J^\alpha\left(\frac{\phi_s\psi_p}{\beta^{-1}}\right) &=&
\int_{x_{\alpha-1}}^{x_\alpha}
\frac{\phi_s(x)\,\psi_p(x)}{\beta^{-1}} \,dx,\label{Jbeta-1}
\end{eqnarray}
$\alpha=1,\dots,n$ ($n$ is the number of layers).

Eq.~(\ref{M12gyr-planar}) accounts for no mode conversion at
transversal magnetization, in agreement with the long ago
established result~\cite{GilliesHlawiczka76}. A few more implicit
tips could also be deduced thereof:

    1. In a symmetrically sandwiched stack, the \textit{polar}
    magnetization $(\mathbf u = u_x\mathbf{\hat x})$ virtually doesn't
    couple the modes exhibiting similar symmetry in the $\phi_s(x)$
    and $\psi_p(x)$ functions (that is, when both $\phi_s(x)$
    and $\psi_p(x)$ are either even or odd); on the contrary, the
    modes of opposite symmetry in $\phi_s(x)$ and $\psi_p(x)$ are
    virtually not sensitive to the \textit{longitudinal}
    magnetization $(\mathbf u = u_z\mathbf{\hat z})$. Asymmetric
    sandwiching impairs this behavior.

    2. The integral (\ref{Jbeta-1}) contains scaling parameter
    $\beta^{-1}$ that defines characteristic thickness of the
    waveguide corresponding to the comparable values of the both
    integrals, Eqs.~(\ref{Jdx}) and (\ref{Jbeta-1}), in Eq.~(\ref{M12gyr-planar}),
    and hence to the comparable shares of polar and longitudinal
    polarizations in the mode conversion efficiency. For optical
    frequencies $\beta^{-1} \simeq 100$ nm; if the thickness of the
    waveguide considerably exceeds $\beta ^{-1}$, than
    $\psi_p(x)$would appear to be a too slowly varying function
    and therefore
    \begin{equation}
    J^\alpha\left(\frac{\phi_s\,d\psi_p}{dx}\right) \ll
    J^\alpha\left(\frac{\phi_s\psi_p}{\beta^{-1}}\right).
    \end{equation}
    For the ultrathin layers the inverse inequality holds, but in
    this regime at most one mode is guided in the structure, which
    is apparently out of the scope here.

\subsection{Optically-inhomogeneous dielectric waveguide}
If we assume $\xi=\xi(\mathbf r)$ in Eq.~(\ref{M12}) to be a
scalar function within the waveguide volume $V$, which conveys the
cases of optical inhomogeneity and ``isotropic'' nonlinearity,
then it is advantageous to simplify Eq.~(\ref{M12}) using the
identity
\begin{equation}\label{VectorIdentity-2}
\bm\nabla\times(\xi\bm\nabla\times\mathbf H) =
\bm\nabla(\bm\nabla\xi\cdot\mathbf H) - (\mathbf H\cdot\bm\nabla)
\bm\nabla\xi
- (\bm\nabla\xi\cdot\bm\nabla)\mathbf H  -
\xi\bm\nabla^2\mathbf H,
\end{equation}
where $\bm\nabla\cdot\mathbf H = 0$ is taken into account. In
planar geometry, we are concerned as before with the TE and TM
modes, but now expressed exclusively through the magnetic field
vectors. For the TM ($p$-polarized) mode we can readily apply
Eq.~(\ref{Hp}); for the TE ($s$-polarized) mode we have for the
electric field, in parallel with Eq.~(\ref{Ds}),
\begin{equation} \label{Es}
\mathbf E_{s} = e^{i\beta z}\varphi_{s}(x) \mathbf{\hat y},
\end{equation}
hence
\begin{equation} \label{Hs}
\mathbf H_{s} = -ik^{-1}\bm\nabla\times\mathbf E_{s} =
k^{-1}e^{i\beta z}\left( \beta\varphi_s(x)\mathbf{\hat x} -
i\frac{d\varphi_s(x)}{dx}\mathbf{\hat z} \right),
\end{equation}
where we once again encounter $\beta$ (in fact, $\beta^{-1}$) in
the role of scaling parameter. If we put, say, $\mathbf H =
\mathbf H_s$ in Eq.~(\ref{VectorIdentity-2}) and multiply the
result by $\mathbf H_p^{\,*}$, then only the first and second
terms of the sum (\ref{VectorIdentity-2}) will produce nonzero
results, as it immediately follows from Eqs.~(\ref{Hs}) and
(\ref{Hp}). Equation (\ref{M12}) thus gives
\begin{equation}\label{M12-diel}
\mathcal{M}_{sp}^{inh} = S^{-1}\!\!\int_W
[\bm\nabla(\bm\nabla\xi\cdot\mathbf H_s)]\cdot\mathbf H_p^{*} \,
dv - S^{-1}\!\!\int_W [(\mathbf H_s\cdot\bm\nabla)\bm\nabla\xi]
\cdot\mathbf H_p^{*} \, dv.
\end{equation}
Now care should be taken when dealing with $\bm\nabla\xi$ function
which is discontinuous in the vicinity of material boundaries
[otherwise --- if $\bm\nabla\xi$ is assumed to be all-continuous
--- Eq.~(\ref{M12-diel}) gives $\mathcal{M}_{sp}^{inh} = 0$]. The safest
way is to introduce a new vector field $\bm\varsigma =
\bm\varsigma(\mathbf r)$ instead of $\bm\nabla\xi$, with a
requirement $\bm\nabla\times\bm\varsigma = 0$ everywhere except at
the boundaries:
\begin{equation}
\bm\nabla\times\bm\varsigma = -(\mathbf{\hat n}\times\bm\varsigma)
\, \delta(\mathbf r - \mathbf R),
\end{equation}
and with this in mind to perform the brute-force evaluation of
integrals in Eq.~(\ref{M12-diel}):
\begin{eqnarray}
\int_W [\bm\nabla(\bm\nabla\xi\cdot\mathbf H_s)]\cdot\mathbf
H_p^{*} \, dv = \beta k^{-1}\!\!
\int_W\frac{\partial\varsigma_x}{\partial y}\varphi_s(x)\psi_p(x)
\, dv - ik^{-1}\!\!\int_W\frac{\partial\varsigma_z}{\partial
y}\frac{d\varphi_s(x)}{dx}\psi_p(x) \, dv,
\\
\int_W [(\mathbf H_s\cdot\bm\nabla)\bm\nabla\xi] \cdot\mathbf
H_p^{*} \, dv = \beta k^{-1}\!\!
\int_W\frac{\partial\varsigma_y}{\partial x}\varphi_s(x)\psi_p(x)
\, dv - ik^{-1}\!\!\int_W\frac{\partial\varsigma_y}{\partial
z}\frac{d\varphi_s(x)}{dx}\psi_p(x) \, dv,
\end{eqnarray}
hence
\begin{eqnarray}\label{M12diel-planar}
\mathcal{M}_{sp}^{inh} &=& \beta(kS)^{-1}\!\!\int_W(-\bm\nabla
\times\bm\varsigma)_z\varphi_s(x)\psi_p(x) \, dv +
i(kS)^{-1}\!\!\int_W(-\bm\nabla\times\bm\varsigma)_x\frac{d\varphi_s(x)}{dx}\psi_p(x)
\, dv
\nonumber\\
&=& \beta(kS)^{-1}\!\!\int_\Sigma\varsigma_y(\mathbf
r)\varphi_s(x)\psi_p(x) \, d\sigma
\nonumber\\
&=& \beta(kS)^{-1}\!\!\int_V \left(
\xi_y'\frac{d(\varphi_s\psi_p)}{dx} +
\xi_{xy}''\varphi_s\psi_p\right) \, dv .
\end{eqnarray}
We see that for the nonzero mode conversion, the $\xi(\mathbf r)$
function should exhibit explicit dependence on the $y$ coordinate.
For this reason $\mathcal{M}_{sp}^{inh}$ would depend on the
transversal dimension $Y$ of the waveguide and on the values of
$\xi(\mathbf r)$ at the $y = 0$ and $y = Y$ boundaries. Another
important consequence relates the effect of the $z$ modulation of
$\xi(\mathbf r)$ on the coupling length. If $\xi(\mathbf r)$ does
not depend on $z$, then $\mathcal{M}_{sp}^{inh}$ would in turn be
independent on the waveguide length $Z$; but on the contrary,
whenever $\xi(\mathbf r)$ is a stochastically oscillating or
periodic function of $z$, the coupling matrix
$\mathcal{M}_{sp}^{inh}$ and hence the coupling length $z_c$ would
become $Z$-dependent. Say, for the $\xi(z)\propto\cos(z/z_0)$
dependence we have $\mathcal{M}_{sp}^{inh}(Z) \propto
Z^{-1}\sin(Z/z_0)$ which is of the order of $Z^{-1}$ and brings
about an unexpected scaling rule $z_c \sim Z$ for the coupling
length in a planar waveguide with periodic modulation of
$\delta\epsilon(\mathbf r)$ in the $z$ direction.

Finally, I would like to remind here that conventional expression
for the off-diagonal coupling matrix element given by the standard
coupled-mode theory~\cite{CoupledModeBOOKs,Huang94},
\begin{equation}
\mathbb{M}_{12} \propto \int\mathbf E_1\,\delta\epsilon\,\mathbf
E_2^{*}\,dx\,dy,
\end{equation}
gives an identical zero for the
$\text{TE}\leftrightarrow\text{TM}$ mode conversion, being the
dielectric perturbation $\varepsilon\,\delta\epsilon(\mathbf r)$ a
scalar function.

\section{\label{sec4}Conclusion}
Mode conversion displays breaking the (initial) symmetry of the
Hamiltonian of a system by some perturbation that distorts the
mode spectrum, i.e., shifts the eigenvalues and alters
polarization of the eigenmodes, hence a natural tool to treat the
mode coupling and conversion phenomena is the perturbation
technique. For the mode conversion at anticrossings, I presented
in this article an electromagnetic counterpart of
quantum-mechanical perturbation theory for quasi-degenerate levels
based on the Lindstedt-Poincar\'{e} method of strained parameters
as a specific alternative to the entrenched coupled-mode formalism
grounded upon the method of variation of independent coefficients.

The general expressions derived for the minimum mode mismatch at
anticrossings, Eq.~(\ref{MinModeMismatch}), and for the coupling
length, Eq.~(\ref{CouplingLength}), are compact, transparent and
premise on the calculation of just one element of the coupling
matrix, Eq.~(\ref{M12}). That matrix element was calculated
explicitly for the two cases of interest in integrated optics: for
a multilayer gyrotropic waveguide under piecewise-constant,
arbitrarily oriented magnetization [Eq.~(\ref{M12gyr-planar})],
and for an optically-inhomogeneous planar dielectric waveguide
[Eq.~(\ref{M12diel-planar})]. In a similar way a large variety of
optical systems can be analyzed, including Bragg's fibers,
photonic crystals with broken periodicity, etc.

Finally, I should note that by means of an appropriate coordinate
mapping, the problem of perturbation due to shifted material
boundaries --- so to say, geometrical perturbation known to spur
difficulties when treated via conventional perturbation
techniques~\cite{Boundaries} --- can be reduced to the problem of
perturbation in permittivity and permeability profiles of a
waveguide exhibiting perfect geometry in those (curvilinear)
coordinates and governed by ``Cartesian-looking'' Maxwell's
equations. Such a trick that recently aided the FDTD modeling of
high index-contrast photonic crystals \cite{WardPendryVsShyroki}
would also significantly augment both the standard coupled-mode
theory and the stationary perturbation technique described here.

\begin{acknowledgments}
I cordially thank Prof. I.~D.~Feranchuk who gave an initial push
for this work by making me look at the whole variety of
quantum-mechanical phenomena associated with quasi-degenerate
levels.
\end{acknowledgments}


\end{document}